\documentclass{iau}
\usepackage{graphicx}

\title[Pulsation-driven mass loss] 
{Pulsation and Mass Loss Across the HR Diagram: From OB stars to Cepheids to Red Supergiants}

\author[Hilding R.~Neilson]   
{Hilding R.~Neilson$^1$
}

\affiliation{$^1$Dept.~of Physics \& Astronomy, East Tennessee State University, , PO Box 70300, Johnson City, TN 37614, USA  email: {\tt neilsonh@etsu.edu} \\[\affilskip]
}
\pubyear{2013}
\volume{301}  
\pagerange{???--???}
\jname{Precision Asteroseismology: Celebration of the Scientific Opus of Wojtek Dziembowski}
\editors{W.~Chaplin, J.~Guzik, G.~Hander \& A.~Pigulski, eds.}

\begin{document}

\maketitle

\begin{abstract}
Both pulsation and mass loss are commonly observed in stars and are important ingredients for understanding stellar evolution and structure, especially for massive stars. There is a growing body of evidence that pulsation can also drive and enhance mass loss in massive stars and that pulsation-driven mass loss is important for stellar evolution. In this review, I will discuss recent advances in understanding pulsation driven mass loss in massive main sequence stars, classical Cepheids and red supergiants and present some challenges remaining.
\keywords{circumstellar matter, stars: evolution, stars: mass loss, Cepheids}
\end{abstract}

\firstsection 
\section{Introduction}

Stellar winds are ubiquitous in massive stars, whether they be hot O-type stars or cool red supergiants and are an crucial ingredient for stellar and galactic evolution. However, the underlying physics of stellar winds and mass loss is still not well-understood across the Hertzsprung-Russell diagram (HRD).

Mass loss in massive stars plays an important role in determining their evolution and how they end as supernovae. The most obvious example of this evolution a Wolf-Rayet stars that eject their envelopes, exposing their helium cores. Because of that mass loss, these stars appear to evolve at hot effective temperatures, $\ge 30,000$~K, and explode as Type I supernovae.  \cite[Georgy (2012)]{Georgy2012} showed that changing the mass-loss rates of red supergiant stars forces them to evolve blue ward and potentially explode during a yellow supergiant phase of evolution, consistent with observations (\cite[Maund et al. 2011]{Maund2011}). Similarly, understanding the wind-driving mechanism and mass-loss rates of red supergiant stars would provide insights into questions about the progenitor masses of Type IIP and IIn supernovae (\cite[Smartt 2009]{Smartt2009}).

Similarly, understanding stellar mass loss provides insight into the circumstellar medium of massive stars and feedback into the interstellar medium.  Evidence for this is found across the HRD, such as bow shocks observed about the O-type star $\zeta$ Oph (\cite[Gvaramadze et al. 2012]{Gvaramadze2012}), the Galactic center star IRS 8 (\cite[Rauch et al. 2013]{Rauch2013}), the prototype Cepheid, $\delta$ Cep (\cite[Marengo et al. 2010]{Marengo2010}) and the red supergiant Betelgeuse (\cite[Ueta et al. 2008; Cox et al. 2012]{Ueta2008, Cox2012}). Further, changes in the wind due to stellar evolution could lead to the formation of multiple bow shocks (\cite[Mohamed et al. 2012; Mackey et al. 2012; Decin et al. 2012]{Mohamed2012, Mackey2012, Decin2012}). Observations of stellar wind bow shocks can be used to infer mass-loss rates and wind velocities.

The challenge for understanding stellar winds is that there are many driving mechanisms possible.  In Fig.~1, I show a HRD outlining regions for different mass-loss mechanisms and regions where mass-loss rates are significant but  not understood, such as in Cepheids, luminous blue variable stars and red supergiant stars.  In hot stars, $T_{\rm{eff}} > 20,000~$K, radiative-line driving is the dominate mechanism for accelerating a stellar wind (\cite[Lamers \& Cassinelli 1999; Vink2011]{Lamers1999, Vink2011}).  In these stars, radiation accelerates Fe III and IV ions and these ions interact with other atoms in the photosphere driving the wind.  In the coolest stars, the mechanism is similar, but instead of accelerating ions, radiation accelerates dust that forms in the stellar photosphere (e.g., \cite[Mattsson et al. 2010; Mattsson \& H\"{o}ffner 2011]{Mattsson2010, Mattsson2011}).  For lower mass stars, like the Sun and red giant stars, winds are driven by Alfenic waves and/or turbulence generating a chromosphere and corona that also accelerates a wind (e.g., \cite[Cranmer \& Saar 2011]{Cranmer2011}).  Rotation is yet another mechanism that can drive mass loss, especially in Be stars (\cite[Porter \& Rivinius 2003]{Porter2003}).

\begin{figure}[t]
\begin{center}
\includegraphics[width=0.95\textwidth]{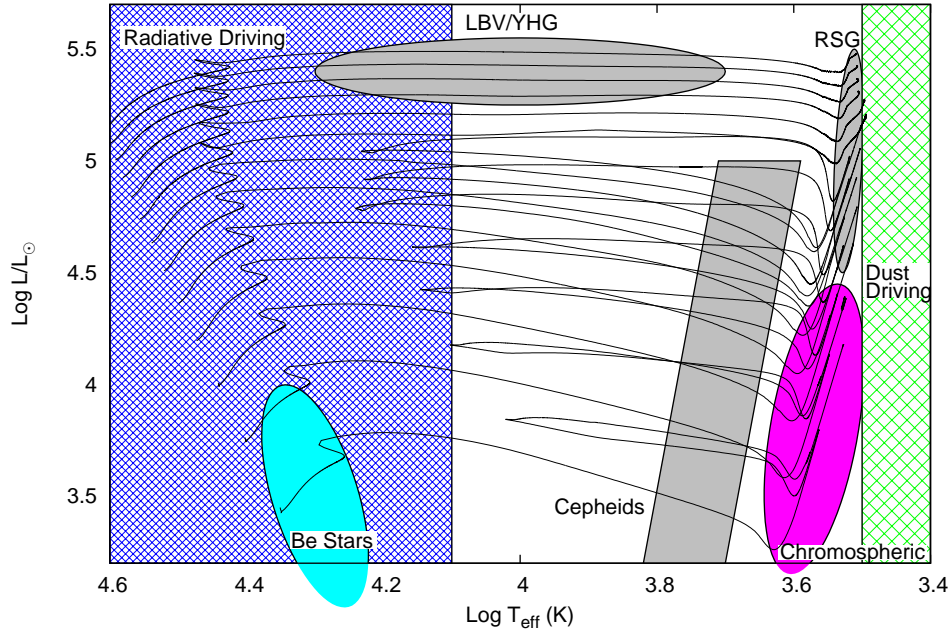}
\end{center}
\caption{Hertzsprung-Russell diagram showing regions of different wind driving mechanisms. Regions denoting Cepheids, luminous blue variables and red supergiants the wind driving mechanism is poorly understood.}
\end{figure}

Even with a plethora of diverse wind-driving mechanisms, there still exist stars for which stellar mass loss is not understood.   However, a potential mechanism for driving mass loss is stellar pulsation, especially in stars such as luminous blue variable stars, Cepheids and red supergiant stars.  In this review, I will discuss progress in understanding how pulsation can drive and influence stellar winds.  However, research on the topic of pulsation-driven mass loss has focused on specific stellar types and not the physics of the pulsation-driven mass loss in general.  As such, I will present recent results as a tour of stars on the HRD where pulsation-driven mass loss is important.

\section{Massive OB stars}
The first step in the tour of the HRD is massive OB stars, in particular OB supergiants.  Radiative-line driving is important, but there is evidence that radiative line-driven theory is insufficient to explain observed mass-loss rates in many of these stars.  \cite[Aerts et al. (2010)]{Aerts2010} presented photometric and spectroscopic observations of the hot supergiant HD 50064
and found variation of P~Cygni line profiles consistent with periodic mass loss.  The period of the mass-loss variation is the same as the period of brightness variations measured from photometry.  The authors suggested that the mass loss must be driven by or is modulated by pulsation, in this case strange mode oscillations (\cite[Glatzel \& Kiriakidis 1993]{Glatzel1993}).  Similarly, Kraus et al. (these proceedings) show similar line profile variations in OB supergiants consistent with pulsation-driven mass loss.  This adds further evidence for the connection between pulsation and mass loss in massive stars, however, these works do not discuss how pulsation can drive a wind. 

\cite[Sanyal \& Langer (2013)]{Sanyal2013} presented preliminary stellar evolution models of a massive star that develops an inflated envelope with a density inversion near the surface.  The model also unstable to pulsation where the radius varies by about 10\%, but the luminosity amplitude is only about ten milli-magnitudes.  This variation is consistent with strange-mode oscillations.  The authors also find that the pulsation instability is connected to the mass loss, but more research is required.

Massive blue supergiants may also pulsate due to the $\epsilon$-mechanism. \cite[Moravveji et al. (2012)]{Moravveji2012} presented evidence for this by modeling the blue supergiant Rigel and finding that the $\epsilon$-mechanism drives gravity-mode oscillations.  It is possible that these gravity modes might contribute to the stellar wind and enhance the mass-loss rate.

\section{Luminous Blue Variable stars}
Luminous blue variables (LBV) stars may be the prototype for pulsation-driven and eruptive mass loss and can be subdivided into two groups: the $\eta$ Carinae-like LBVs and the S Doradus stars. The former stars are primarily eruptive variables, named for $\eta$ Car which is famous for observed eruptions in the 19th and early 20th-century (\cite[Humphreys \& Koppelman 2005; Davidson \& Humphreys 2012]{Humphreys2005, Davidson2012}). The latter stars are defined by more regular pulsation and microvariations (\cite[e.g., Saio et al. (2013)]{Saio2013}).

The cause of the observed eruptions in $\eta$ Car are still a mystery. \cite[Smith (2013)]{Smith2013} modeled the great 19th century eruption as a strong wind with mass-loss rates, $\dot{M} = 0.33~M_\odot~$yr$^{-1}$ lasting 30 years followed by an explosion, ejecting about 10~$M_\odot$ of material.  The model is similar to that expected for a Type IIn supernova. However, the source of the explosion was not described.  \cite[Guzik et al. (2005)]{Guzik2005} computed hydrodynamic models of massive stars that undergo pulsation and found that pulsation could  cause the radiative luminosity in the star to increase with a delay before the convective luminosity compensates.  The radiative luminosity surpasses the Eddington luminosity, hence driving an eruption. In this scenario, pulsation in stars near the Eddington limit is a potential mass-loss mechanism.

The S Doradus stars appear to pulsate more regularly than the $\eta$ Carinae stars where pulsation is driven by either the iron bump opacity or the strange mode instability.  The stars also straddle the effective temperature, $T_{\rm{eff}} \approx 22,000~$K, which defines the boundary of the bi-stability mechanism for radiative-driven winds (\cite[e.g., Vink et al. 2013]{Vink2013}).  For $T_{\rm{eff}} > 22,000~$K, the dominant opacity for driving the wind is the Fe IV ions, while for cooler stars the dominant opacity is Fe III.  The change in opacity causes significant differences in mass-loss rates, cooler stars having greater mass-loss rates than on the hotter side of the bi-stability region.  Pulsation could drive a star to oscillate between the hot and cool side of the S Doradus instability strip and change the mass-loss rate (\cite[Smith et al. 2004]{Smith2004}).  
Pulsation can also directly drive mass loss in these stars, possibly by strange mode oscillations  (\cite[Grott et al. 2005]{Grott2005}) .  A combination of the two mechanisms might generate a pseudo-photosphere, making a S Dor star appear as a yellow hypergiant, as suggested by \cite[Vink 2012]{Vink2012}.  As before, stellar pulsation can enhance the wind, but the underlying physics is uncertain.

\section{$\beta$ Cephei stars}
$\beta$ Cephei stars are massive stars where pulsation is driven by the iron opacity bump that oscillate both radially and non-radially with pulsation periods of hours.  \cite[Guzik \& Lovekin (2012)]{Guzik2012} speculated that $\beta$ Cephei stars may lose mass via the same super-Eddington luminosity events as in LBV stars. However, these stars appear to have the opposite problem as LBV stars; the observed mass-loss rates are smaller than expected from radiative line-driven theory. \cite[Oskinova et al. 2011]{Oskinova2011} presented observations of seven $\beta$ Cephei stars plus a number of non-pulsating, magnetic B-type stars, all of which have smaller mass-loss rates than expected, i.e., the weak-wind problem (e.g. \cite[Crowther et al. 2006]{Crowther2006}).

One phenomenon common to all these stars is X-ray emission, which affects the ionization structure in the photosphere, hence changing the observed mass-loss rate.  Classical Cepheids also display X-ray emission that is correlated with pulsation (\cite[Engle \& Guinan 2012]{Engle2012}); this hints at the possibility that X-ray emission in $\beta$ Cep stars is also pulsation related.  There is no evidence for phase-dependent X-ray emission (\cite[Raassen et al. 2005]{Raassen2005}) but the pulsation period is short relative to the integration time required to detect X-ray photons.  If pulsation is generating X-ray emission then pulsation can be acting to stall the stellar wind.

\section{Be stars}
The Be stars are an interesting class of stars displaying emission lines consistent with the presence of a circumstellar mass loss disk generated by rapid rotation (\cite[Struve 1931]{Struve1931}).  However, it  is not obvious how a disk is generated if Be stars rotate at less than their critical velocity (e.g., \cite[Marsh Boyer et al. 2012]{Marsh2012}), although this is still an issue of contention (\cite[Rivinius 2013]{Rivinius2013}).  One potential resolution is stellar pulsation and pulsation-driven mass loss (\cite[Baade 1983; Ando 1986]{Baade1983, Ando1986}).

Be stars are well-known non-radial pulsating stars, where gravity modes have been observed (\cite[Diago et al. 2009; Neiner et al. 2009; Semaan et al. 2013]{Diago2009, Neiner2009, Semaan2013}).  Pulsation provides a number of potential avenues to help generate mass loss, either by pulsation-rotation interactions (e.g., Townsend, these proceedings; Goupil, these proceedings), or directly as suggested by \cite[Cranmer (2009)]{Cranmer2009} and Shibahashi (these proceedings). 

For the latter case, \cite[Cranmer (2009)]{Cranmer2009} presented an analytic prescription for driving a disk outflow in Be stars with rotation velocities as slow as 60\% of the critical velocity, where non-radial pulsations transfer angular momentum and accelerates that material to a keplerian velocity. This model is somewhat ad hoc as there is no obvious mechanism for transferring the angular momentum.  However, Shibashi does build upon models of wave leakage in the upper photosphere (e.g.,  \cite[Townsend 2000a,b]{Townsend2000a, Townsend2000b}) providing a plausible connection. The \cite[Cranmer (2009)]{Cranmer2009} mass-loss mechanism depends on non-radial pulsations instead of radial pulsation, but still suffers the same issue of not being able to detail the connection between pulsation and mass loss beyond some prescription.

\section{Classical Cepheids}
From the hot star side of the HRD, I consider cooler effective temperatures with a discussion of   Classical Cepheids.  These stars are arguably the archetype for stellar pulsation; they are bright radially-pulsating stars that are ideal standard candles for cosmology (\cite[Freedman et al. 2001; Ngeow et al. 2009; Freedman et al. 2012]{Freedman2001, Ngeow2009, Freedman2012}). However, recent infrared and radio observations show that the prototypical Cepheid, $\delta$ Cephei, is undergoing mass loss (\cite[Marengo et al. 2010; Matthews et al. 2012]{Marengo2010, Matthews2012}).  This result was surprising and raises important questions about the role of pulsation and mass loss in these stars.

There is evidence for mass loss in other Cepheids based on infrared observations.  \cite[Barmby et al. (2011)]{Barmby2011} presented Spitzer infrared observations of Galactic Cepheids and found, at best, tentative evidence for infrared excess in these stars.  On the other hand, \cite[Neilson et al. 2009, 2010]{Neilson2009b, Neilson2010} analyzed optical and infrared observations of Large Magellanic Cloud Cepheids and found evidence that infrared excess is common and suggesting mass-loss rates of the order $\dot{M} = 10^{-8}$ - $10^{-7}~M_\odot~$yr$^{-1}$. 

However, infrared excess is not the only indicator for Cepheid mass loss. \cite[Neilson et al. (2012a)]{Neilson2012a} showed that the observed rate of period change and other fundamental parameters  of the Cepheid Polaris cannot be fit by stellar evolution models unless Polaris is losing mass at a rate of the order $10^{-7}$ - $10^{-6}~M_\odot~$yr$^{-1}$.  This result was extended to a population of almost 200 Galactic Cepheids (\cite[Turner et al. 2006]{Turner2006}).  For that case, \cite[Neilson et al. (2012b)]{Neilson2012b} compared the observed fraction of Cepheids with positive and negative period change with predictions from population synthesis models.  A positive rate of period change is consistent with a Cepheid evolving to hotter effective temperatures and a negative rate is consistent with red ward evolution.  Mass loss acts to increase the rate of period change in a positive direction, but it also decreases the timescale of blue ward evolution.  The observed fraction of Cepheids with positive period change is about 67\%, and \cite[Neilson et al. (2012b)]{Neilson2012b} predicted the fraction is more than 80\% if mass loss is not included.  When stellar evolution models are computed with Cepheid mass-loss rates, $\dot{M} = 10^{-7}~M_\odot$~yr$^{-1}$ then the predicted fraction is reduced to about 70\%, consistent with observations.  This is the first evidence that all Cepheids undergo significant mass loss.

While mass loss is important in Cepheids, there is no obvious mass-loss theory. \cite[Neilson \& Lester (2008, 2009)]{Neilson2008, Neilson2009a} developed a prescription for pulsation-driven mass loss in Cepheids, in which a wind is driven by pulsation-generated shocks.  Previous calculations suggest there are multiple shocks propagating in a Cepheid photosphere at various phases  (\cite[Fokin et al. 1996]{Fokin1996}) and it was hypothesized that these shocks add momentum to an outflow, similar to that suggested by \cite[Bowen (1988)]{Bowen1988}.  Mass-loss rates for a sample of Galactic Cepheids were predicted to be $\dot{M} = 10^{-10}$ - $10^{-7}~M_\odot~$yr$^{-1}$, smaller than those suggested by infrared observations and rates of period change.  However, when the theory was incorporated with stellar evolution models, the mass-loss rates were sufficient to resolve the Cepheid mass discrepancy (\cite[Keller 2008; Neilson et al. 2011]{Keller2008, Neilson2011}). The pulsation-driven mass-loss prescription is interesting, but insufficient to describe observations.

\section{Red Supergiant stars}
A Cepheid evolves to cooler effective temperatures and becomes a red supergiant (RSG) star.  Mass loss in RSGs is typically understood in terms of the Reimer's relation or a similar type of mass-loss prescription (e.g., \cite[Reimers et al. 1975; Schr{\"o}der \& Cuntz 2005]{Reimers1975, Schroder2005}). However, the Reimer's relation is a measure of average mass-loss rates and RSG winds can vary significantly (\cite[Willson 2000]{Willson2000}). While mass-loss rates have been measured, the driving mechanism is not understood even though mass loss is a crucial ingredient for stellar evolution and supernova progenitors (\cite[Langer 2012]{Langer2012}).

Radial pulsation was suggested as one mechanism for driving super winds in RSG stars. \cite[Yoon \& Cantiello (2010)]{Yoon2010} found that RSG stellar evolution models to be pulsationally unstable near the end of their lives.  Pulsation amplitudes were found to increase and the authors assumed that this pulsation drives a super wind.  Mass-loss rates were assumed to be a function of the amplitude growth rate. The enhanced mass-loss rates are significant enough for RSG stars to lose three or more solar masses, hence enabled \cite[Yoon \& Cantiello (2010)]{Yoon2010} to hypothesize that pulsation-driven mass loss could explain the observed dearth of Type IIP supernovae progenitors with mass $ M \ge 16.5~M_\odot$, as noted by \cite[Smartt (2009)]{Smartt2009}. 

\cite[Georgy (2012)]{Georgy2012} also computed stellar evolution models with enhanced mass loss.  He found that by increasing mass-loss rates by factors of three to five would cause a RSG star to evolve blue ward to hotter effective temperatures and end up as yellow supergiants when they explode as supernovae.  This model also assumes that RSG stars undergo pulsation-driven mass loss, but there is no defined theory and almost no observational evidence for pulsation-driven mass loss.

However, the observed bow shock structure of Betelgeuse provides a tantalizing hint for pulsation-driven mass loss.  \cite[Decin et al. (2012)]{Decin2012} found multiple arc structures about Betelgeuse's bow shock. These multiple arcs are too close to be a result of stellar evolution, but may be caused by pulsation changing the structure of the stellar wind.  This hypothesis is speculative and needs to be tested by hydrodynamic models.   Further work is required to begin to understand the role of pulsation-driven mass loss in RSG stars.

\section{AGB stars}
Asymptotic giant branch stars are perhaps the best understood examples of stars undergoing pulsation-driven mass loss.  These stars are particularly cool, allowing dust to form in the photosphere.  Radiation then accelerates the dust in a wind, analogous to radiative-driven winds in hot stars.  However, this dust-driven wind is efficient only in the coolest and/or carbon-rich stars (e.g., \cite[Wachter et al. 2008]{Wachter2008}). In other AGB stars, such as the M-type AGB stars, dust does not form easily, hence another wind-driving mechanism is necessary (\cite[Woitke 2006]{Woitke2006}).  

Stellar pulsation is the most likely candidate for driving mass loss in these stars.  \cite[H{\"o}fner \& Anderson 2007]{Hofner2007} suggested that pulsation in M- and S-type AGB stars generate shocks that levitate material, extending the photosphere.  The extended, cooler photosphere allows for silicate dust to form that would not form otherwise.  Radiation can then take over and accelerate a wind. \cite[Freytag \& H{\"o}fner (2008)]{Freytag2008} computed three-dimensional simulations of an AGB atmosphere that included convection and pulsation and verified the hypothesis that pulsation could levitate material in the photosphere.  Furthermore, \cite[Bladh \& H{\"o}fner (2011)]{Bladh2012} and \cite[Bladh et al. (2013)]{Bladh2013} presented radiation-hydrodynamic atmosphere models, again,  showing that the combination of the dust formation in the photosphere and pulsation drives an outflow. Currently, understanding AGB mass loss is limited more by knowledge of what dust species form in the photosphere than the role of pulsation.

\section{Outlook}

In this review, I described how stellar pulsation can influence mass loss in a number of different stellar types ranging  from massive OB stars to coolest AGB stars. There has been significant progress in understanding pulsation-driven mass loss in the past decade, including some of the first observational hints.   

While observations are beginning to hint at pulsation-driven mass loss, there is not yet a proverbial smoking gun. The most compelling observations include spectral line variations in massive OB stars and the large mass-loss rates measured for Cepheids.  For the most part, evidence for pulsation-driven mass loss is the inability for other driving mechanisms to explain observed mass-loss rates, even in LBV and AGB stars. An ideal observation connecting pulsation and mass loss would be variations of P~Cygni profiles in Cepheids and AGB stars.  In hot stars, other observations are necessary to confirm pulsation-driven mass loss since P~Cygni profiles measure mass-loss rates from radiative line driving. Variations of those profiles measure changes of mass-loss rates and/or changes due to pulsation velocities.  

Observational evidence is still circumstantial partly because there is also no general theory for pulsation-driven mass loss.  Current theories are targeted toward specific stellar types and specific challenges, such as explosive events in $\eta$ Carinae LBV stars and dust formation in AGB stars.  For instance, the theory of radiative-line driven winds is general (\cite[Castor et al 1975]{Castor1975}) and can be applied to all stars; that theory is only limited to hot stars because of how it depends on the ionization structure of the photosphere.  The one distinct challenge for developing a general theory for pulsation-driven mass loss is, in most cases, pulsation is not the sole driving mechanism.  In hot stars line driving is the dominant mechanism while in cool stars dust driving is the dominant mechanism.  Pulsation enhances mass loss in these stars, thus any pulsation-driven mass-loss theory must couple with other wind driving mechanisms.  

Developing a theory of pulsation-driven mass loss is a difficult challenge, but the future is bright. Hydrodynamic modeling, such as that by \cite[Guzik \& Lovekin (2012)]{Guzik2012} and \cite[Bladh et al. (2013)]{Bladh2013}  is progressing and new insights are being discovered.  New models will shed further light on mass loss in LBV and AGB stars as well as explore the physics for pulsation and mass loss in $\beta$ Cephei, RSG, massive blue supergiant, and Be stars as well as Classical Cepheids.  Greater insight in pulsation-driven mass loss will also provide deeper understanding of the detailed stellar evolution and precision stellar astrophysics.



\end{document}